\documentclass[10pt,aps,prd,floatfix,notitlepage,showpacs,twocolumn]{revtex4-1}

\usepackage[utf8x]{inputenc}

\usepackage{amsmath,amssymb,amsthm}
\usepackage{comment}
\usepackage{multirow}
\usepackage{graphicx}

\begin{document}

\title{Generalized hybrid metric-Palatini gravity}

\author{N.~Tamanini and C.~G.~B\"ohmer}
\affiliation{Department of Mathematics, University College London, Gower Street, London, WC1E 6BT, UK}
\email{n.tamanini.11@ucl.ac.uk}
\email{c.boehmer@ucl.ac.uk}
\date{\today}

\begin{abstract}
We introduce a new approach to modified gravity which generalizes the recently proposed hybrid metric-Palatini gravity. The gravitational action is taken to depend on a general function of both the metric and Palatini curvature scalars. The dynamical equivalence with a non-minimally coupled bi-scalar field gravitational theory is proved. The evolution of cosmological solutions is studied using dynamical systems techniques.
\end{abstract}

\maketitle

\section{Introduction}

In order to explain the present and initial accelerated expansions of the Universe a large variety of modified theories of gravity has been proposed in recent years. Among them, one of the most popular is $f(R)$ modified gravity where the gravitational action depends on a general function of the curvature scalar $R$, see \cite{Sotiriou:2008rp} for two reviews. To derive the gravitational field equations from those modified actions, two approaches are extensively used in the literature: the metric and the Palatini variational principles (See \cite{Amendola:2010bk} for recent extensions of the Palatini variational method in modify gravity). In the so called metric approach one takes the metric $g_{\mu\nu}$ as the only dynamical variable and considers only variations of the action with respect to it. The so called Palatini approach is based on the idea of considering the connection defining the Riemann curvature tensor to be {\it a priori} independent of the metric. As such, one performs variations of the action with respect to the metric and the connection independently. Both approaches have been used extensively to build cosmological models, many of which contain an era of accelerated expansion.

It is well-known that $f(R)$ theories are dynamically equivalent to Brans-Dicke (BD) theories. In fact, metric $f(R)$ gravity has been shown to be dynamically equivalent to a Brans-Dicke theories with vanishing BD parameter while Palatini $f(R)$ gravity presents the same equivalence if the BD parameter equals $-3/2$ (see again \cite{Sotiriou:2008rp}). The value $-3/2$ for the BD parameter is a peculiar one since it implies no dynamics for the scalar field in BD theories. Consequently, Palatini $f(R)$ gravity has the same number of dynamical degrees of freedom as general relativity, see~\cite{Boehmer:2013ss} for a very different model which does not introduce new dynamical degrees of freedom either. 

More recently, a novel approach to modified gravity has been introduced where a Palatini-like $f(\mathcal R)$ term is added to the metric Einstein-Hilbert action \cite{Harko:2011nh}. In this context cosmological and astrophysical applications together with wormholes geometries have been studied in \cite{Capozziello:2012ny}, where it has also been shown that viable accelerating cosmological solutions are allowed by some specific models. The theory is dynamically equivalent to a scalar-tensor theory with non-minimal coupling to gravity given by $(1+\phi) R$ where $R$ is the metric curvature scalar and $\phi$ is the scalar field. This is done in strict analogy with Palatini $f(R)$ gravity and indeed the BD parameter for this theory is still $-3/2$. However, because of the different coupling to $R$ than in BD theories, in this theory the scalar field is dynamical and represents a new dynamical degree of freedom.

In the present paper we analyze a natural extension of the theory introduced in \cite{Harko:2011nh}. We introduce a general function which depends on both the metric and Palatini curvature scalars. We show that this new generalization can be considered as dynamically equivalent to a gravitational theory with two scalar fields. Only one of these scalar fields is non-minimally coupled to $R$ and in general an interaction between the two appears in the action. The cosmological features of the theory are then studied using dynamical system techniques.

Let us define $\kappa^2=8\pi G/c^4$ and start from the action
\begin{align}
  S_{f}=\frac{1}{2\kappa^2}\int d^4x \sqrt{-g}\, f(R,\mathcal{R}) \,,
  \label{004}
\end{align}
which generalizes the so-called hybrid metric-Palatini action of \cite{Harko:2011nh}. In action (\ref{004}) $R$ is the Ricci curvature scalar formed with the Levi-Civita connection,
\begin{align}
  \Gamma^\lambda_{\mu\nu} = \frac{1}{2} g^{\lambda\sigma} \left(\partial_\mu g_{\nu\sigma}+\partial_\nu g_{\mu\sigma}-\partial_\sigma g_{\mu\nu}\right) \,,
\end{align}
while $\mathcal R$ is the curvature scalar of an independent torsionless connection $\hat\Gamma^\lambda_{\mu\nu}$, in analogy with the Palatini approach.
The variation of the action~(\ref{004}) with respect to the independent connection $\hat\Gamma^\lambda_{\mu\nu}$ leads to
\begin{align}
  \hat\nabla_\lambda\left(\sqrt{-g}\frac{\partial f}{\partial \mathcal{R}} g^{\mu\nu}\right)=0 \,,
\end{align}
whose solution is a Levi-Civita connection in terms of the conformal metric $h_{\mu\nu}=\frac{\partial f}{\partial \mathcal{R}}g_{\mu\nu}$,
\begin{align}
  \hat\Gamma^\lambda_{\mu\nu} = \frac{1}{2} h^{\lambda\sigma} \left(\partial_\mu h_{\nu\sigma}+\partial_\nu h_{\mu\sigma}-\partial_\sigma h_{\mu\nu}\right) \,.
  \label{001}
\end{align}

The variation with respect to the metric yields
\begin{multline}
  \frac{\partial f}{\partial{R}} R_{\mu\nu} -\frac{1}{2}g_{\mu\nu}f \\
  -\left(\nabla_\mu\nabla_\nu-g_{\mu\nu}\Box \right)\frac{\partial f}{\partial{R}}  +\frac{\partial f}{\partial \mathcal{R}}\mathcal{R}_{\mu\nu}=8\pi T_{\mu\nu} \,,
  \label{002}
\end{multline}
where also the matter action has been considered in the variation. Because of (\ref{001}), $\mathcal{R}_{\mu\nu}$ can be related to $R_{\mu\nu}$ and terms involving (derivatives of) $\partial f/\partial\mathcal{R}$ as in Palatini $f(R)$. However the trace of (\ref{002}) now relates $\mathcal{R}$ and its derivatives to $T$ and $g_{\mu\nu}$, meaning that it is not possible in general to solve for $\mathcal{R}$. However, this can be avoided by requiring $\partial^2f/\partial{R}\partial{\mathcal{R}}=0$. In this particular case the trace of (\ref{002}) becomes an algebraic equation in $\mathcal{R}$ which can be solved for $T$ and (derivatives of) $g_{\mu\nu}$. Note that in the following we will keep the function $f$ completely arbitrary as it turns out that all results we will obtain hold for any (sufficiently smooth) function $f$.

\section{Dynamically equivalent actions and conformal transformations}

Let us start by considering the action
\begin{multline}
  S=\frac{1}{2\kappa^2}\int d^4x\sqrt{-g} \Bigg[f(\alpha,\beta)+\frac{\partial f(\alpha,\beta)}{\partial\alpha}\left(R-\alpha\right) \\
    +\frac{\partial f(\alpha,\beta)}{\partial\beta} \left(\mathcal{R}-\beta\right)\Bigg] \,,
  \label{003}
\end{multline}
where $\alpha$ and $\beta$ are two scalar fields. Variation with respect to $\alpha$ and $\beta$ gives the system
\begin{align}
  \frac{\partial^2f}{\partial\alpha^2}\left(R-\alpha\right) +\frac{\partial^2f}{\partial\alpha\partial\beta}\left(\mathcal{R}-\beta\right) =0 \,,\\
  \frac{\partial^2f}{\partial\alpha\partial\beta}\left(R-\alpha\right) +\frac{\partial^2f}{\partial\beta^2}\left(\mathcal{R}-\beta\right) =0 \,,
\end{align}
whose only solution is given by $\alpha=R$ and $\beta=\mathcal{R}$, provided that
\begin{align}
  \frac{\partial^2 f}{\partial\alpha^2}\frac{\partial^2 f}{\partial\beta^2} \neq \left(\frac{\partial^2 f}{\partial\alpha\partial\beta}\right)^2 \,.
  \label{029}
\end{align}
This condition follows simply from requiring that this matrix type equation is non-degenerate. It is interesting to note that the matrix involved is in fact the Hessian of $f$. Since our theory is based on an action principle, a non-degenerate Hessian in this context means nothing but that solutions of the field equations derived from the action are indeed stationary points of the action. 

It is now clear that substituting this solution back into action (\ref{003}) produce immediately action (\ref{004}). The two actions are thus dynamically equivalent.
Constraint (\ref{029}) excludes from our analysis the cases when the function $f$ is linear in either $\alpha$ (i.e.~$R$) or $\beta$ (i.e.~$\mathcal R$). However the first case is nothing but the hybrid metric-Palatini theory studied in \cite{Harko:2011nh,Capozziello:2012ny}, while the second is equivalent to usual metric $f(R)$ theories. Moreover constraint (\ref{029}) excludes also some particular models such as $f=\exp(R+\mathcal{R})$ or $f=\sqrt{R\,\mathcal R}$. We have thus to reduce the results of this section to the models satisfying (\ref{029}).

Let us define two new scalar fields as
\begin{align}
  \chi=\frac{\partial f(\alpha,\beta)}{\partial\alpha} \quad\mbox{and}\quad \xi=-\frac{\partial f(\alpha,\beta)}{\partial\beta} \,.
\end{align}
The minus sign in the definition of $\xi$ is required in order not to allow for a negative kinetic energy of the field.
Action (\ref{003}) can be rewritten as
\begin{align}
  S=\frac{1}{2\kappa^2}\int d^4x\sqrt{-g} \left[\chi R-\xi \mathcal{R}- V(\chi,\xi)\right] \,,
  \label{040}
\end{align}
where the interaction potential is defined as
\begin{align}
  V(\chi,\xi)=-f(\alpha(\chi),\beta(\xi))+\chi\,\alpha(\chi) -\xi\,\beta(\xi) \,.
\end{align}
Due to solution (\ref{001}) (which can also be obtained varying action (\ref{040}) with respect to $\hat\Gamma^\lambda_{\mu\nu}$) we can expand $\mathcal{R}$ and find (up to boundary terms)
\begin{align}
  S=\frac{1}{2\kappa^2}\int d^4x\sqrt{-g} \left[\left(\chi-\xi\right)R-\frac{3}{2\xi}(\partial\xi)^2 -V(\chi,\xi)\right] \,.
\end{align}
We can shift $\chi$ by $\xi$ defining a new scalar field as $\phi=\chi-\xi$. In this way the action becomes
\begin{align}
  S=\frac{1}{2\kappa^2}\int d^4x\sqrt{-g} \left[\phi\,R-\frac{3}{2\xi
    }(\partial\xi)^2 -W(\phi,\xi)\right] \,.
  \label{030}
\end{align}
It is possible to think of action (\ref{030}) as a Brans-Dicke theory with vanishing BD parameter and a potential interacting with another minimally coupled scalar field.

At this point we can perform a conformal transformation in order to switch from the Jordan to the Einstein frame. The transformation
\begin{align}
  g_{\mu\nu}\mapsto \tilde g_{\mu\nu}=\phi\, g_{\mu\nu} \,,
\end{align}
allows then, up to surface terms, to rewrite action (\ref{030}) as
\begin{multline}
  S=\frac{1}{2\kappa^2}\int d^4x\sqrt{-\tilde g} \Bigg[\tilde R -\frac{3}{2\phi^2
    }(\partial\phi)^2 \\
    -\frac{3}{2\phi\xi}(\partial\xi)^2 -\frac{W(\phi,\xi)}{\phi^2}\Bigg] \,.
\end{multline}
Finally, we redefine the two scalar fields as
\begin{align}
  \tilde\phi = \sqrt{\frac{3}{2}} \frac{\ln\phi}{\kappa} \quad\mbox{and}\quad \tilde\xi = \frac{2\sqrt{2}}{\kappa}\sqrt{\xi} \,,
\end{align}
and the action becomes
\begin{multline}
  S=\int d^4x \sqrt{-\tilde g} \Bigg[\frac{1}{2\kappa^2} \tilde R -\frac{1}{2} (\tilde\nabla\tilde\phi)^2 \\
    -\frac{1}{2} e^{-\sqrt{2}\kappa\tilde\phi/\sqrt{3}} (\tilde\nabla\tilde\xi)^2 -\tilde W(\tilde\phi,\tilde\xi) \Bigg] \,,
  \label{031}
\end{multline}
where the new potential is defined as
\begin{align}
  \tilde W(\tilde\phi,\tilde\xi) =\frac{1}{2\kappa^2} e^{-2\sqrt{2}\kappa\tilde\phi/\sqrt{3}} W(e^{\sqrt{2}\kappa\tilde\phi/\sqrt{3}},\kappa^2\tilde\xi^2/8) \,.
\end{align}

Action (\ref{031}) is well-known within the context of the so-called Brans-Dicke or two-field inflation \cite{Berkin:1991nm,GarciaBellido:1995fz,Cid:1900zz}, where the scalar field $\phi$ represents the Brans-Dicke field while $\xi$ denotes the inflaton. Usually these studies start from a more general action than (\ref{030}) where also a kinetic term for $\phi$ is considered, but do not allow for a coupling between the two scalar fields in the potential \cite{Berkin:1991nm}. A more general coupling between the scalar $\phi$ and $R$ was considered in \cite{GarciaBellido:1995fz}. In the Einstein frame this leads to a general function of $\tilde\phi$ in the exponential coupling to the kinetic term of $\tilde\xi$. In general we can address action (\ref{031}) as a specific model of Brans-Dicke inflation with vanishing BD parameter. This means that from hybrid metric-Palatini gravity we have a natural explanation for introducing both the inflaton and the Brans-Dicke scalar fields. We refer to \cite{Cid:1900zz} for recent developments in the context of two-field inflation.

In the next section we will analyze the general cosmological dynamics of the theory in the Einstein frame.

\section{Cosmological dynamics in the Einstein frame}

For the sake of simplicity, from now on we will omit the tildas in action (\ref{031}). In other words, in what follows we will denote with $g_{\mu\nu}$, $\phi$, $\xi$ and $W$ the quantities in the Einstein frame. The field equations can be obtained by varying action (\ref{031}) with respect to the dynamical variables $g_{\mu\nu}$, $\phi$ and $\xi$. The gravitational field equations in the Einstein frame are thus given by
\begin{align}
  G_{\mu\nu}=\kappa^2 \left(T_{\mu\nu}^{(\phi)} +e^{-\kappa\phi\sqrt{2/3}}\,T_{\mu\nu}^{(\xi)}-g_{\mu\nu}W\right) \,,
  \label{032}
\end{align}
where we define
\begin{align}
  T_{\mu\nu}^{(\phi)}&=\nabla_\mu\phi\nabla_\nu\phi-\frac{1}{2}g_{\mu\nu} (\nabla\phi)^2 \,,\\
  T_{\mu\nu}^{(\xi)}&=\nabla_\mu\xi\nabla_\nu\xi-\frac{1}{2}g_{\mu\nu} (\nabla\xi)^2 \,.
\end{align}
For the moment we assume that every other form of matter is negligible in comparison to the two scalar fields. The equations for the two scalar fields are given by
\begin{align}
  \Box\phi +\frac{\kappa}{\sqrt{6}} e^{-\kappa\phi\sqrt{2/3}} (\nabla\xi)^2 -W_\phi &=0 \,, \label{033}\\
  \Box\xi +\frac{\kappa\sqrt{2}}{\sqrt{3}} \nabla_\mu\phi\nabla^\mu\xi -e^{\kappa\phi\sqrt{2/3}}\,W_\xi &= 0 \,,\label{034}
\end{align}
where $W_\phi$ and $W_\xi$ are the derivatives of the potential with respect to $\phi$ and $\xi$ respectively.

We consider a cosmological FLRW metric
\begin{align}
ds^2=-dt^2+a(t)^2\left(\frac{dr^2}{1-k r^2}+r^2d\Omega^2\right) \,,
\end{align}
where $a(t)$ is the scale factor.
From the gravitational field equations (\ref{032}) we obtain the following cosmological equations
\begin{align}
3\frac{k}{a^2}+3H^2= \frac{\kappa^2}{2}e^{-\sqrt{2/3} \kappa  \phi} \dot\xi^2+\frac{\kappa^2}{2}\dot\phi^2+ \kappa^2\,W \,,\label{036}\\
\frac{k}{a^2}+2 \dot H+3 H^2=-\frac{\kappa^2}{2}e^{-\sqrt{2/3} \kappa  \phi} \dot\xi^2 -\frac{\kappa^2}{2}\dot\phi^2+\kappa^2\,W \,. \label{037}
\end{align}
The scalar fields equations (\ref{033}) and (\ref{034}) give the following evolution equations
\begin{align}
\ddot\phi+3 H \dot\phi+\frac{\kappa}{\sqrt{6}}  e^{-\sqrt{2/3} \kappa  \phi}\, \dot\xi^2+W_\phi=0 \,,\label{038}\\
\ddot\xi+3 H \dot\xi-\frac{\kappa\sqrt{2}}{\sqrt{3}}\, \dot\xi\,\dot\phi+e^{\sqrt{2/3} \kappa  \phi}\, W_\xi=0 \,.\label{039}
\end{align}

In what follows we will consider only spatially flat ($k=0$) cosmological models. In order to recast the cosmological equations (\ref{036})--(\ref{039}) into a dynamical system, we will make use of the following dimensionless variable 
\begin{align}
  x^2= \frac{\kappa^2\dot\phi^2}{6H^2}\,, \quad y^2=\frac{\kappa^2 W}{3H^2}\,, \quad s^2=\frac{\kappa^2 \dot\xi^2}{6H^2}e^{-\sqrt{2/3} \kappa  \phi}\,.
\label{041}
\end{align}
The definitions of the $x$ and $y$ variables have been extensively considered to study the cosmological dynamics in both uncoupled and coupled dark energy-dark matter models \cite{Wetterich:1994bg}. With (\ref{041}) the Friedmann constraint (\ref{036}) reads
\begin{align}
  x^2+y^2=1-s^2 \,,
\end{align}
impliying that
\begin{align}
  0\leq x^2+y^2 \leq 1 \,,
\end{align}
since $0\leq s^2\leq 1$. Moreover because of the positiveness of the potential energy we must have $y\geq 0$, which implies that $x$ and $y$ can only take values within half a unit disc.

In order to complete the autonomous system of equations coming from the cosmological equations (\ref{037})--(\ref{039}) we must specify the potential $W$. In the following we will consider three possible cases for the potential energy and will analyse the outcoming phase spaces.

\subsection{Model 1: $W=W_0\,e^{-\lambda\kappa\phi/\sqrt{6}}$}

\begin{table*}
\caption{Critical points and their properties for model 1.}
\label{tab:01}
\begin{tabular}{|c|c|c|c|c|c|c|}
\hline
Point & x & y & Existence & $w_{eff}$ & Acceleration & Stability \\
\hline
\hline
$A_-$ & -1 & 0 & $\forall\;\lambda$ & 1 & No & Saddle \\
\hline
\multirow{2}*{$A_+$} & \multirow{2}*{1} & \multirow{2}*{0} & \multirow{2}*{$\forall\;\lambda$} & \multirow{2}*{1} & \multirow{2}*{No} & Unstable if $\lambda\leq 6$\\ & & & & & & Saddle if $\lambda>6$ \\
\hline
\multirow{2}*{$B$} & \multirow{2}*{$\frac{6}{\lambda+2}$} & \multirow{2}*{$\frac{\sqrt{2}}{\sqrt{\lambda+2}}$} & \multirow{2}*{$\lambda\geq\sqrt{37}-1$} & \multirow{2}*{$\frac{\lambda-2}{\lambda+2}$} & \multirow{2}*{No} & Saddle if $\lambda=\sqrt{37}-1$ \\ & & & & & & Stable spiral if $\lambda>\sqrt{37}-1$ \\
\hline
\multirow{2}*{$C$} & \multirow{2}*{$\frac{\lambda}{6}$} & \multirow{2}*{$\frac{\sqrt{36-\lambda^2}}{6}$} & \multirow{2}*{$\lambda\leq$ 6} & \multirow{2}*{$\frac{\lambda^2}{18}-1$} & \multirow{2}*{$\lambda<2\sqrt{3}$} & Stable if $\lambda<\sqrt{37}-1$ \\ & & & & & & Saddle if $\sqrt{37}-1\leq\lambda\leq 6$ \\
\hline
\end{tabular}
\end{table*}

First we will consider the usual quintessence exponential potential given by
\begin{align}
W=W_0\,e^{-\lambda\kappa\phi/\sqrt{6}} \,,
\label{048}
\end{align}
where $W_0$ and $\lambda$ are both positive constants.
In this model the only interaction between $\phi$ and $\xi$ is given by the kinetic coupling being the potential $\phi$ dependent only. The scalar field $\phi$ plays the role of the usual quintessence dark energy field, while we can consider $\xi$ as representing dark matter. Both dark matter and dark energy have thus a geometrical origin in this model and no particles have to be introduced opportunely.

In terms of the variables (\ref{041}) equation (\ref{037}) becomes
\begin{align}
\frac{\dot H}{H^2} = 3\left(y^2-1\right) \,,
\label{042}
\end{align}
which always gives a scaling solution for $a(t)$ in terms of the value of $y$. If $y=1$ we have $\dot H=0$ and the universe undergoes an exponential expansion, while if $y>\sqrt{2/3}$ the universe undergoes a scaling accelerated expansion. From (\ref{042}) we can read off the effective equation of state parameter of the total energy content of the universe as
\begin{align}
w_{eff} = 1-2y^2 \,.
\label{044}
\end{align}

The autonomous system of equation is in this case two dimensional and it is given by equations (\ref{038}) and (\ref{039}) as
\begin{align}
x'&= x^2-3\, x\, y^2+\frac{1}{2} (\lambda +2)\, y^2-1 \,,\\
y'&= -\frac{1}{2}\, y \left(\lambda \, x+6 y^2-6\right) \,,
\end{align}
where a prime denotes differentiation with respect to $N=\ln a$. There are up to four critical points for this system. The points and their properties are showed in Table \ref{tab:01}. There are three possible qualitative behaviours of the phase space depending on the following three range for $\lambda$: $0<\lambda\leq\sqrt{37}-1$, $\sqrt{37}-1\leq\lambda\leq 6$ and $\lambda>6$.

\begin{figure}
\centering
\includegraphics[width=\columnwidth]{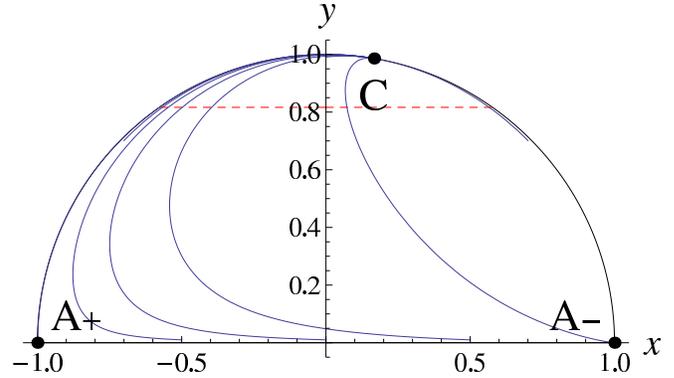}
\caption{Phase space for Model 1 with $\lambda=1$. The global attractor represents an accelerating solution because it lies in the region above the dashed/red line.}
\label{fig:01}
\end{figure}

\begin{figure}
\centering
\includegraphics[width=\columnwidth]{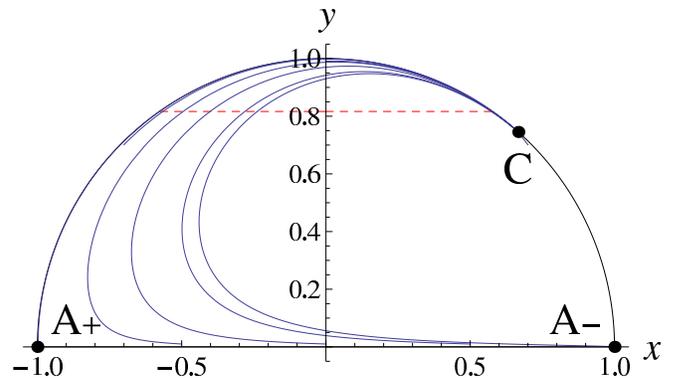}
\caption{Phase space for Model 1 with $\lambda=4$. The global attractor does not represent an accelerating solution because it lies in the region below the dashed/red line.}
\label{fig:02}
\end{figure}

From Table \ref{tab:01} we see that in order to have a stable accelerated attractor we must have $\lambda<2\sqrt{3}$, so the most interesting solutions will belong to the first range whose phase space is showed in Fig.~\ref{fig:01}. Points $A_-$ and $A_+$ are respectively a saddle and unstable point and represent early time solutions with a stiff fluid effective equation of state. Every solution evolves eventually reaching point $C$ which always lies on the unit circle and acts as a global attractor. If $\lambda<2\sqrt{3}$ point $C$ belongs to the region above the dashed/red line where the universe is accelerating. If instead $2\sqrt{3}\leq\lambda\leq\sqrt{37}-1$ point $C$ will be below the dashed/red line and the universe will end in a decelerating solution. However we can still have a period of accelerated expansion because, for a wide range of initial conditions, the evolution still pass through the accelerated region for some time as is showed in Fig.~\ref{fig:02}. If $\lambda=3\sqrt{2}$ the global attractor will represent a matter dominated universe with vanishing effective equation of state parameter, while for $\lambda=2\sqrt{6}$ the final state will be a radiation-like dominated universe which suggests that this model could be of interest in early time inflationary dynamics.

\begin{figure}
\centering
\includegraphics[width=\columnwidth]{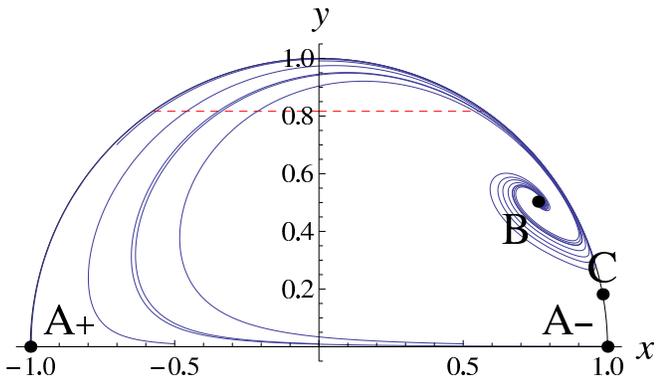}
\caption{Phase space for Model 1 with $\lambda=5.9$. The global attractor is now point $B$, while $C$ is a saddle point.}
\label{fig:03}
\end{figure}

The phase space for the range $\sqrt{37}-1\leq\lambda\leq 6$ has been drawn in Fig.~\ref{fig:03}. Point $B$ is now the global attractor and always represents a decelerating solution since can only appear below the dashed/red line. Point $C$ is now a saddle point which attracts all the trajectories before they turn to point $B$. Again, depending on the initial conditions, the universe can still undergo a phase of accelerated expansion since several trajectories pass through the accelerated region above the dashed/red line.

\begin{figure}
\centering
\includegraphics[width=\columnwidth]{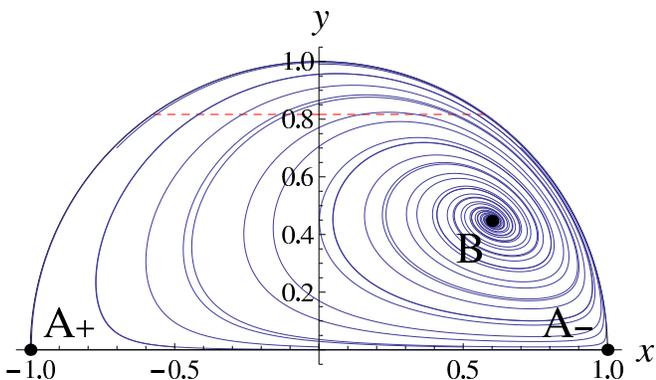}
\caption{Phase space for Model 1 with $\lambda=8$. The global attractor is point $B$, while now $A_+$ is a saddle point.}
\label{fig:04}
\end{figure}

The last range for which the dynamics of Model 1 is different is given for $\lambda>6$. Its phase space is depicted in Fig.~\ref{fig:04}. Point $A_+$ is now a saddle point and attracts all the trajectories along the $y$ direction. The global attractor is still point $B$ and the cosmological evolution can experience more than one eras of accelerated expansion before ending eventually in the final decelerating solution.

In conclusion we have seen that in Model 1 the universe can undergoes phases of accelerated expansion for all the possible range of $\lambda$. If $\lambda<2\sqrt{3}$ the cosmic evolution will end in an accelerated state, while for all the other values of $\lambda$ it eventually reaches a stable decelerating solution.

\subsection{Model 2: $W = \alpha \, (\kappa\,\xi)^\lambda \, e^{-\lambda \kappa \phi /\sqrt{6}}$}

\begin{table}
\caption{Critical points and their properties for model 2.}
\label{tab:02}
\begin{tabular}{|c|c|c|c|c|c|c|}
\hline
Point & x & y & z & Existence & $w_{eff}$ & Acceleration \\
\hline
\hline
$A_-$ & -1 & 0 & 0 & $\forall\;\lambda$ & 1 & No \\
\hline
\multirow{2}*{$A_+$} & \multirow{2}*{1} & \multirow{2}*{0}& \multirow{2}*{0} & \multirow{2}*{$\forall\;\lambda$} & \multirow{2}*{1} & \multirow{2}*{No} \\ & & & & & & \\
\hline
$B_-$ & -1 & 0 & 1 & $\forall\;\lambda$ & 1 & No \\
\hline
$B_+$ & 1 & 0 & 1 & $\forall\;\lambda$ & 1 & No \\
\hline
\multirow{2}*{$C_0$} & \multirow{2}*{$\frac{6}{\lambda+2}$} & \multirow{2}*{$\frac{\sqrt{2}}{\sqrt{\lambda+2}}$} & \multirow{2}*{0} & \multirow{2}*{$\lambda\geq\sqrt{37}-1$} & \multirow{2}*{$\frac{\lambda-2}{\lambda+2}$} & \multirow{2}*{No} \\ & & & & & & \\
\hline
\multirow{3}*{$D_0$} & \multirow{3}*{$\frac{\lambda}{6}$} & \multirow{3}*{$\frac{\sqrt{36-\lambda^2}}{6}$} & \multirow{3}*{0} & \multirow{3}*{$\lambda\leq$ 6} & \multirow{3}*{$\frac{\lambda^2}{18}-1$} & \multirow{3}*{$\lambda<2\sqrt{3}$} \\ & & & & & & \\ & & & & & & \\
\hline
\multirow{3}*{$D_1$} & \multirow{3}*{$\frac{\lambda}{6}$} & \multirow{3}*{$\frac{\sqrt{36-\lambda^2}}{6}$} & \multirow{3}*{1} & \multirow{3}*{$\lambda\leq$ 6} & \multirow{3}*{$\frac{\lambda^2}{18}-1$} & \multirow{3}*{$\lambda<2\sqrt{3}$} \\ & & & & & & \\ & & & & & & \\
\hline
\end{tabular}
\end{table}

\begin{table*}
\caption{Critical points and their stability properties for model 2.}
\label{tab:02a}
\begin{tabular}{|c|c|c|c|c|c|c|c|}
\hline
Point & Eigenvalues & Stability \\
\hline
\hline
$A_-$ & $3$, $-2$, $3 + \lambda/2$ & Saddle point \\
\hline
$A_+$ & $3$, $2$, $3 - \lambda/2$ & Saddle point \\
\hline
$B_-$ & $-3$, $-2$, $3 + \lambda/2$ & Saddle point\\
\hline
$B_+$ & $-3$, $2$, $3 - \lambda/2$ &  Saddle point\\
\hline
$C_0$ & $3\lambda/(2 + \lambda)$, $(-3 \pm \sqrt{81 + 32 \lambda - 4 \lambda^2 - \lambda^3})/(2 + \lambda)$ & Saddle point \\
\hline
$D_0$ & $\lambda^2/12$, $-3 + \lambda^2/12$, $-6 + \lambda/3 + \lambda^2/12$  & Saddle point \\
\hline
\multirow{2}*{$D_1$} & \multirow{2}*{$-\lambda^2/12$, $-9/2 + \lambda/6 + \lambda^2/8$, $-\infty$} & Stable if $\lambda < 2(\sqrt{82}-1)/3$ \\
& & `Saddle' if $ (2\sqrt{82}-2)/3 < \lambda$\\
\hline
\end{tabular}
\end{table*}

In this section we will consider the potential given by
\begin{align}
  W(\phi,\xi) = \alpha \, (\kappa\,\xi)^\lambda \, e^{-\lambda \kappa \phi /\sqrt{6}} \,,
  \label{035}
\end{align}
where $\alpha$ and $\lambda$ are two dimensionless positive parameters. This potential allows for a direct coupling between the two scalar fields $\phi$ and $\xi$. Unfortunately, we cannot recast the cosmological evolution equations (\ref{037})--(\ref{039}) into a two dimensional dynamical system. However defining in addition to (\ref{041}) the new variable
\begin{align}
  z=\frac{H_0}{H+H_0} \,,
  \label{043}
\end{align}
we can obtain a three dimensional system. The new variable $z$ has been choosen in such a way to mantain the phase space compact~\cite{Boehmer:2008av}. It takes values between $0$ and $1$, meaning that the phase space is now represented by a half cilinder with radius and heigth equal to one. Equations (\ref{037}) can still be rewritten as (\ref{042}) implying that at every point of the phase space the effective equation of state is again given by (\ref{044}). The accelerated region is now the part of the half cilinder corresponding to $y>\sqrt{2/3}$ for all the possible values of $z$.

The cosmological equations (\ref{037})--(\ref{039}) give the following three dimensional dynamical system
\begin{align}
  x'&= \frac{y^2}{2} \left(\lambda+2 -6\, x\right)+x^2-1 \,,\label{045}\\
  y'&= y \Bigg[3-3\, y^2-\frac{\lambda}{2}\, x \nonumber\\
    &\qquad\qquad+\beta\, \sqrt{1-x^2-y^2}  \left(\frac{z}{y\,(1-z)}\right)^{2/\lambda }\Bigg] \,,\label{046}\\
  z'&= 3 \left(y^2-1\right) \left(z-1\right) z \,,\label{047}
\end{align}
where again a prime denotes a derivative with respect to $N=\ln a$ and we have redefine the parameter $\alpha$ as
\begin{align}
  \beta=\frac{\lambda}{\sqrt{2}}\,3^{\frac{\lambda-2}{2\lambda}} \,\alpha^{1/\lambda}\,(\frac{\kappa}{H_0})^{2/\lambda}\,.
\end{align}
In the above system of equations, the term with $\beta$ in (\ref{046}) becomes singular as $y \rightarrow 0$ or $z \rightarrow 1$ and one must be rather careful when investigating the equations for those values. The critical points are given in Table \ref{tab:02} and, according to the value of $\lambda$, there can be up to seven critical points. However, when considering the critical points with $z=1$, we have assumed that the term proportional to $\beta$ approaches zero when $z \rightarrow 1$. This issue is very difficult to settle analytically, however, the numerical solutions and the resulting phase space confirm that this assumption is valid. 

Again the three ranges $\lambda<\sqrt{37}-1$, $\sqrt{37}-1\leq\lambda\leq 6$ and $\lambda>6$ gives the three qualitatively different behaviours of the phase space. The four points $A_\pm$ and $B_\pm$ always represent universes evolving with a stiff matter effective equation of state. However they are expected to be relevant only at early time and not to be stable solutions.

\begin{figure}
\centering
\includegraphics[width=\columnwidth]{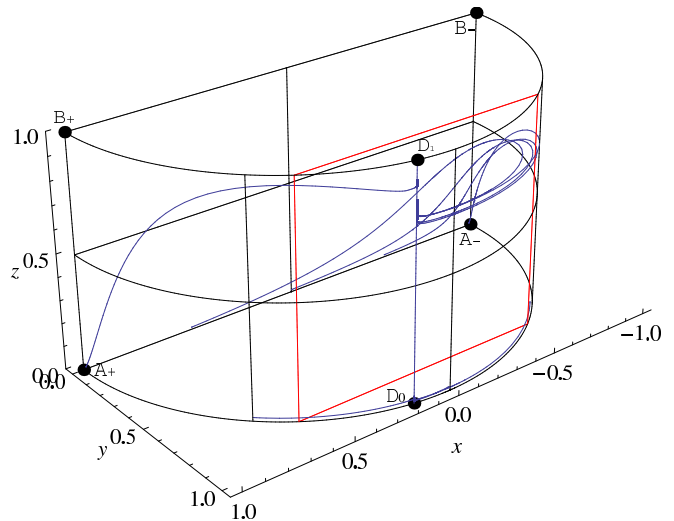}
\caption{Phase space for Model 2 with $\lambda=1$ and $\beta=0.01$. The global attractor is point $D_1$ representing an accelerating solution whenever $\lambda<2\sqrt{3}$ because lies in the region marked by the dashed/red line.}
\label{fig:10}
\end{figure}

The phase space for the first range is shown in Fig.~\ref{fig:10}. Points $A_+$ and $D_0$ act as saddle points attracting the early time solutions before these turn towards greater values of $z$. The trajectories always evolve towards point $D_1$ which represents the global attractor. A few more remarks are required about this point. Firstly, one of the eigenvalues approaches $-\infty$ as $z \rightarrow 1$. The term responsible for this is $\beta\, \sqrt{1-x^2-y^2}  \left(\frac{z}{y\,(1-z)}\right)^{2/\lambda }$, which we assumed to approach zero. The linear stability matrix (Jacobian) contains the inverse of this term which in turn can yield an eigenvalue which formally is $-\infty$. This explains why this points act as the global attractor to the system. There is a $\lambda$ range where one of the eigenvalues of $D_1$ is positive, this point will still attract all trajectories. While there is a direction in which the point repels trajectories, the non-linearities of the system will move any trajectory away from this exact direction and the attractive behaviour in the other directions will dominate. This behaviour can be seen quite clearly in the phase space plots.  

If $\lambda<2\sqrt{3}$ this characterizes an accelerating scaling solution, while if $\lambda> 2\sqrt{3}$ the universes undergoes decelaration. Again this represent the cosmologically interesting case where the global attractor of the phase space could represent an accelerating universe.

\begin{figure}
\centering
\includegraphics[width=\columnwidth]{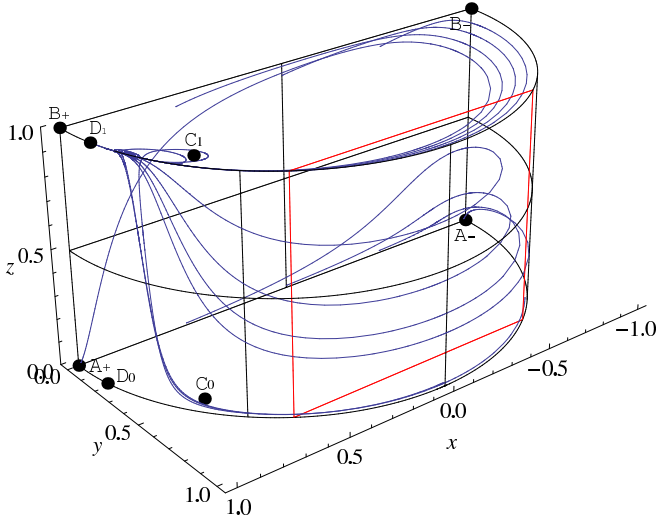}
\caption{Phase space for Model 2 with $\lambda=5.9$ and $\beta=0.01$. The global attractor is still point $D_1$ but now the trajectories are first attracted by point $C_1$ which is not a critical point.}
\label{fig:11}
\end{figure}

The phase space for the second qualitative range $\sqrt{37}-1\leq\lambda\leq 6$ is depicted in Fig.~\ref{fig:11}. Points $A_+$, $C_0$ and $D_0$ represent saddle points which attract the early time solutions. Point $D_1$ still represents the global attractor but do not characterizes an accelerating solution being always outside the accelerated region. 

\begin{figure}
\centering
\includegraphics[width=\columnwidth]{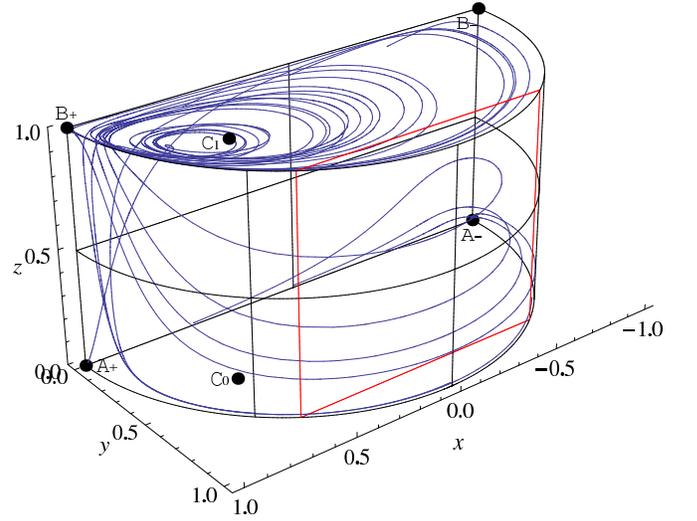}
\caption{Phase space for Model 2 with $\lambda=8$ and $\beta=0.01$.}
\label{fig:12}
\end{figure}

We have identified another interesting point, $C_1$ with coordinates $x=\frac{6}{\lambda+2}$, $y=\frac{\sqrt{2}}{\sqrt{\lambda+2}}$, $z=1$ which is not a critical point to the dynamical system. However, when evaluating equations~(\ref{045}) and~(\ref{047}), we note that both right-hand sides vanish. The remaining equation~(\ref{046}) at this point is
\begin{align}
  y'_{C_1} &= \frac{\beta}{2} \left(\frac{\lambda}{2}+1\right)^{1/\lambda-3/2} \sqrt{\lambda^2 + 2\lambda -36} \left(\frac{z}{1-z}\right)^{2/\lambda } \nonumber \\
  &= \beta C(\lambda) \left(\frac{z}{1-z}\right)^{2/\lambda }\,,
\end{align}
where $C(\lambda)$ is a constant depending on $\lambda$. We can now understand the phase space at this point, see Figure~\ref{fig:12}. If $\beta$ is chosen to be small, then the solution behaves as if $C_1$ was a critical point. The smaller the value of $\beta$ the better $C_1$ acts as an attracting point. However, when $z$ gets sufficiently close to $1$, the trajectories will get repelled from this point eventually. If $\lambda> 6$ none of the critical points is stable, and the solution will keep evolving without a determined late time behaviour. As can be seen in Figure~\ref{fig:12}, all trajectories reach the $z \rightarrow 1$ surface and then stay there. 

\subsection{Model 3: $W=W_0\,e^{-\lambda\kappa\phi/\sqrt{6}}$ + Matter}

\begin{table*}
\caption{Critical points and their properties for model 3.}
\label{tab:03}
\begin{tabular}{|c|c|c|c|c|c|c|c|}
\hline
Point & x & y & z & Existence & $w_{eff}$ & Acceleration & Stability \\
\hline
\hline
$A_-$ & -1 & 0 & 0 & $\forall\;\lambda$ & 1 & No & Saddle \\
\hline
\multirow{2}*{$A_+$} & \multirow{2}*{1} & \multirow{2}*{0}& \multirow{2}*{0} & \multirow{2}*{$\forall\;\lambda$} & \multirow{2}*{1} & \multirow{2}*{No} & Unstable if $\lambda\leq 6$\\ & & & & & & & Saddle if $\lambda>6$ \\
\hline
$B$ & $\frac{6}{\lambda+2}$ & $\frac{\sqrt{2}}{\sqrt{\lambda+2}}$ & 0 & $\lambda\geq\sqrt{37}-1$ & $\frac{\lambda-2}{\lambda+2}$ & No & Saddle \\
\hline
\multirow{2}*{$C$} & \multirow{2}*{$\frac{\lambda}{6}$} & \multirow{2}*{$\frac{\sqrt{36-\lambda^2}}{6}$} & \multirow{2}*{0} & \multirow{2}*{$\lambda\leq$ 6} & \multirow{2}*{$\frac{\lambda^2}{18}-1$} & \multirow{2}*{$\lambda<2\sqrt{3}$} & Stable if $\lambda<3\sqrt{2(w+1)}$ \\ & & & & & & & Saddle otherwise \\
\hline
$D$ & 0 & 0 & 1 & $\forall\;\lambda$ & $w$ & No & Saddle \\
\hline
\multirow{2}*{$E$} & \multirow{2}*{$\frac{3(w+1)}{\lambda}$} & \multirow{2}*{$\frac{3\sqrt{1-w^2}}{\lambda}$} & \multirow{2}*{$\frac{\sqrt{\lambda^2-18(w+1)}}{\lambda}$} & \multirow{2}*{$\lambda\geq 3\sqrt{2(w+1)}$} & \multirow{2}*{$w$} & \multirow{2}*{No} & Stable if $3\sqrt{2(w+1)}\leq\lambda$ \\ 
& & & & & & & Saddle otherwise \\ 
\hline
\end{tabular}
\end{table*}

In this final section we reconsider Model 1 with the potential (\ref{048}) and add a standard matter perfect fluid energy-momentum tensor to the gravitational field equations (\ref{032}),
\begin{align}
G_{\mu\nu}=\kappa^2 \left(T_{\mu\nu}^{(\phi)} +e^{-\kappa\phi\sqrt{2/3}}\,T_{\mu\nu}^{(\xi)} -g_{\mu\nu}W+T_{\mu\nu}^{(M)}\right) \,,
\label{049}
\end{align}
where
\begin{align}
T_{\mu\nu}^{(M)} = p\, g_{\mu\nu} + (p+\rho)\,U_\mu U_\nu \,,
\end{align}
with $U^\mu$ the comoving four-velocity of the fluid, $\rho$ its energy density and $p$ its pressure. We will consider a linear equation of state given by
\begin{align}
p=w\,\rho \,,
\end{align}
and assume that standard matter is covariantly conserved
\begin{align}
\nabla^\mu T_{\mu\nu}^{(M)} = 0 \,.
\label{053}
\end{align}
We will also consider only the physical range $0\leq z \leq 1/3$, meaning that we cannot have cosmic acceleration from matter alone.
In this model dark matter is included in the standard matter sector while both the scalar fields $\phi$ and $\xi$ act as dark energy. Note that we are adding a matter action to (\ref{031}) and thus assuming that the matter fields couple with the Einstein frame metric $\tilde g_{\mu\nu}$. This procedure is different to adding a matter action directly to (\ref{004}) but has largely been considered in literature without entering in deep phylosophical issues (see \cite{Faraoni:2004pi} for a discussion).

The new cosmological field equations derived from (\ref{049}), with $k=0$, read
\begin{align}
3H^2=\kappa^2\rho+ \frac{\kappa^2}{2}e^{-\sqrt{2/3} \kappa  \phi} \dot\xi^2+\frac{\kappa^2}{2}\dot\phi^2+ \kappa^2\,W \,,\label{050}\\
2 \dot H+3 H^2=-\kappa^2p-\frac{\kappa^2}{2}e^{-\sqrt{2/3} \kappa  \phi} \dot\xi^2 -\frac{\kappa^2}{2}\dot\phi^2+\kappa^2\,W \,, \label{051}
\end{align}
while the two evolution equations for the scalar fields still coincide with (\ref{038}) and (\ref{039}). These equations can be recasted in a three dimensional autonomous system of equations defining, in addition to (\ref{041}), the new adimensional variable
\begin{align}
z^2 = \frac{\kappa^2\rho}{3H^2} \,.
\end{align}
The Friedmann constraint (\ref{050}) reduces to
\begin{align}
x^2+y^2+z^2=1-s^2 \,,
\label{052}
\end{align}
implying that the phase space is now the quarter of a unit sphere because, thanks to (\ref{052}) and the positiveness of $\rho$ and $W$, we must have $y\geq 0$, $z\geq 0$ and
\begin{align}
0\leq x^2+y^2+z^2\leq 1 \,.
\end{align}
Equation (\ref{051}) becomes
\begin{align}
\frac{\dot H}{H^2} = \frac{3}{2}\left[-2+2\,y^2+(1-w)\,z^2\right] \,,
\end{align}
from which we can read off the new effective equation of state parameter
\begin{align}
w_{eff} = 1-2\, y^2+(w-1)\, z^2 \,.
\end{align}

\begin{figure}
\centering
\includegraphics[width=\columnwidth]{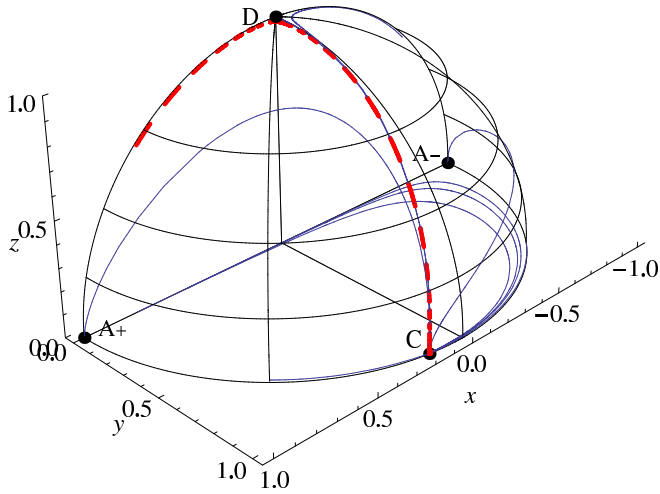}
\caption{Phase space for Model 3 with $\lambda=1$ and $w=0$. The global attractor is point $C$ which represents an accelerating solution when $\lambda<2\sqrt{3}$, while point $D$ is a matter dominated saddle point.}
\label{fig:05}
\end{figure}

\begin{figure}
\centering
\includegraphics[width=\columnwidth]{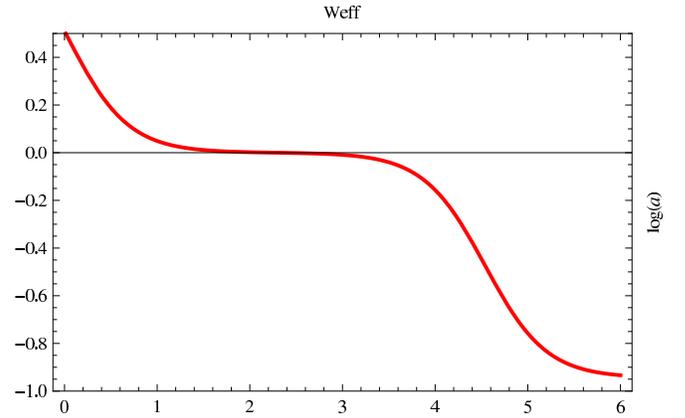}
\caption{Evolution of the effective equation of state parameter for the dashed/red trajectories in Fig.~\ref{fig:05}. After reaching a long lasting matter dominated evolution in according to observations, the universe eventually ends in an accelerated expansion representing the final cosmological stage.}
\label{fig:06}
\end{figure}

The 3D dynamical system is given by equations (\ref{053})--(\ref{051}) as
\begin{align}
x'&=-\frac{3}{2} x \left[2\, y^2-(w-1)\, z^2\right] \nonumber\\
&\qquad\qquad+\frac{1}{2} (\lambda +2)\, y^2+x^2+z^2-1 \,,\\
y'&=-\frac{1}{2} y \left[-3\, (w-1)\, z^2+\lambda\,  x+6\, y^2-6\right] \,,\\
z'&=\frac{3}{2} z \left[\left(w-1\right) \left(z^2-1\right)-2\, y^2\right] \,,
\end{align}
and the critical points together with their properties are listed in Table \ref{tab:03}. If we compare this with Table \ref{tab:01} we see that we have now two more critical points ($D$ and $E$) corresponding to a universe evolving in accordance with the matter equation of state parameter. Point $D$ corresponds to a universe completely dominated by the matter sector with no dark energy affecting the evolution. Point $E$ presents instead both matter and dark energy but the total outcome on the universe evolution is still completely equivalent to a matter dominated universe. The other points, belonging to the $z=0$ plane, have the same properties of Model 1 with point $C$ being the cosmic accelerated stable attractor solution for $\lambda<2\sqrt{3}$.
We have now four qualitative behaviours for the dynamics of the phase space depending again on the possible values of $\lambda$: $\lambda<3\sqrt{2(w+1)}$, $3\sqrt{2(w+1)}\leq\lambda<\sqrt{37}-1$, $\sqrt{37}-1\leq\lambda\leq 6$ and $\lambda>6$. Note that because $0\leq z\leq 1/3$ we always have $3\sqrt{2(w+1)}<\sqrt{37}-1$.

\begin{figure}
\centering
\includegraphics[width=\columnwidth]{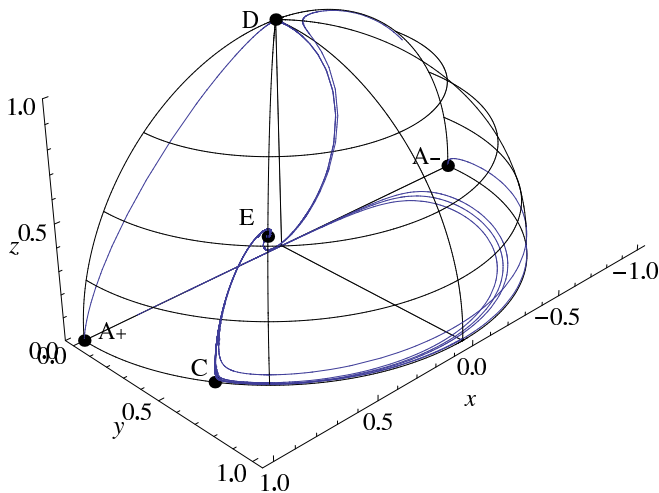}
\caption{Phase space for Model 3 with $\lambda=5$ and $w=0$. The global attractor is now point $E$ where the universe evolves according to the matter equation of state.}
\label{fig:07}
\end{figure}

The first range is again the more interesting since we can have a late time attractor where the universe is accelerating its expansion. The dynamics of the phase space is depicted in Fig.~\ref{fig:05}. The late time attractor is point $C$ which results in an accelerating solution whenever $\lambda<2\sqrt{3}$. Point $D$ represents a saddle point where the universe is completely dominated by the matter sector and expands according to radiation/dust solutions. It is then clear that every trajectory passing nearby point $D$ and eventually ending in point $C$ describes a possible physical universe. In fact all these solutions will allow the universe to undergo the standard radiation and matter eras before the transition to the dark energy accelerating solution.

\begin{figure}
\centering
\includegraphics[width=\columnwidth]{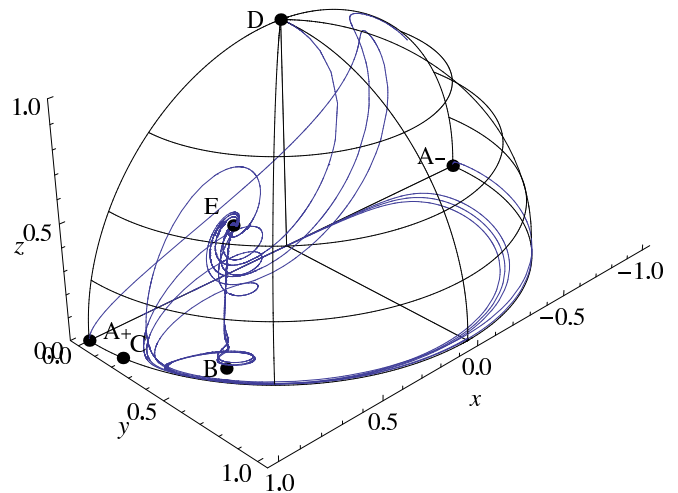}
\caption{Phase space for Model 3 with $\lambda=5.9$ and $w=1/3$. The global attractor is still point $E$ but now the trajectories starting from the $z=0$ plane are first attracted by saddle point $B$.}
\label{fig:08}
\end{figure}

As an example we can look at the dashed/red solution in Fig.~\ref{fig:05} and see how the effective equation of state parameter evolves. This is plotted in Fig.~\ref{fig:06}. We see that the universe immediately reaches a matter dominated expansion and keeps this evolution for some time unperturbed. Of course if the matter equation of state parameter $w$ changed during this period, from radiation to dust in a physical situation, also $w_{eff}$ would change according to $w$. This means that during this period the universe has the time to undergoes the standard cosmological eras in agreement with the observations. Depending on the initial conditions this stage can last for the time needed to produce the nucleosyntesis and create the cosmological structures. Eventually we will have the transition to the accelerated phase which represents the final phase of the universe where the effective equation of state parameter assumes the value $\lambda^2/18-1$. The situation is completely equivalent to Quintessence plus dark matter with $\phi$ playing the role of the dark energy scalar field. In fact the trajectories confined to the border of the sphere, such as the dashe/red one in Fig.~\ref{fig:05}, must have $s=0$ meaning that $\xi$ does not influence the dynamics of this solutions.

\begin{figure}
\centering
\includegraphics[width=\columnwidth]{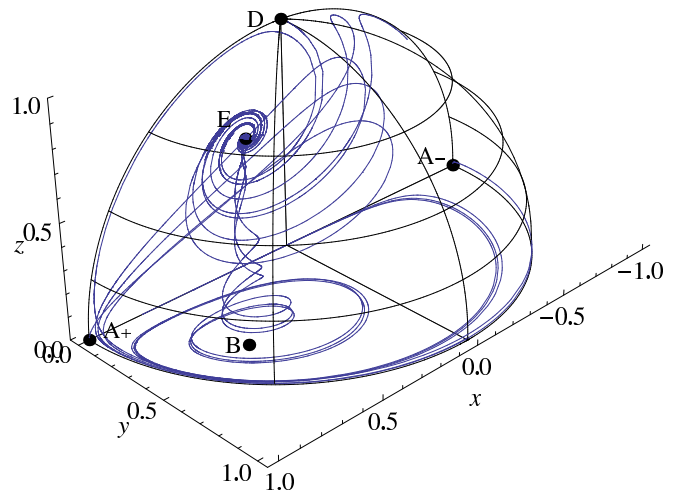}
\caption{Phase space for Model 3 with $\lambda=8$ and $w=1/3$. The global attractor is point $E$ but now $A_+$ is a saddle point attracting the $z=0$ trajectories.}
\label{fig:09}
\end{figure}

We can now look at the second range $3\sqrt{2(w+1)}\leq\lambda<\sqrt{37}-1$ whose phase space is drawn in Fig.\ref{fig:07}. The global attractor is now point $E$ where the universe expands according to the matter equation of state. Depending on the initial conditions the trajectories can either pass nearby saddle point $D$, where we also have $w_{eff}=w$, or pass through the region surrounding the point $(0,1,0)$ where the universe accelerates its expansion and eventually approaches saddle point $C$ before ending to point $E$. This shows how this model can be useful in early inflationary dynamics since we can have an accelerated period before the universe start to be radiation/dust dominated.

This feature is presented also in the third possible range $\sqrt{37}-1\leq\lambda\leq 6$ as Fig.~\ref{fig:08} shows. For this reason and because the phase space properties are more evident, we chose to show the dynamics in the $w=1/3$ case. We still have point $E$ as the global attractor, but now, besides point $C$, also point $B$ act as saddle point influencing trajectories starting from the $z=0$ plane. The dynamics is similar to the second $\lambda$-range with several solutions experiencing an accelerated phase before ending in the matter dominated final evolution.

Finally the phase space for the last range $\lambda>6$ is presented in Fig.~\ref{fig:09}. Point $C$ is now gone and point $A_+$ plays its role attracting the $z=0$ trajectories. The dynamics is again similar to the previous ranges with point $E$ being the global attractor and solutions having a possible era of cosmic accelerated expansion depending on the initial conditions.

\section{Conclusions}

In this paper we have studied a natural generalization of the so-called hybrid metric-Palatini gravity introduced in \cite{Harko:2011nh}. A completely  arbitrary function of both the metric and Palatini curvature scalars was considered as the Lagrangian density in the action. Using dynamically equivalent actions and conformal transformation techniques, we have shown that this new theory can be recast into general relativity plus two scalar fields coupled with each other. Therefore, using this approach one arrives naturally at theories where the different matter components couple to each other. It should be emphasized that there are no theoretical restrictions when it comes to coupling different matter components, all that is required by general relativity and its generalisation is that the total energy-momentum tensor is conserved.   

We analyzed the possible applications to cosmology considering a FLRW universe and employing dynamical system methods. Three specific models specified by their potentials were studied in detail and in each case a late time cosmological accelerated solution has been found. Depending on the model parameters these can represent global attractor solutions. We also encountered a rather peculiar parameter choice (Fig.~\ref{fig:12}) where the dynamical system has no global attractor and the cosmological solution would never stop evolving. 

The first model has a potential without coupling the two scalar fields and shows several similarities with usual quintessence models. The second model considers a direct coupling between the two scalars in the potential and it is characterized by more mathematical complexity. Its evolution is easily understood. Finally, in the third and most interesting model we add standard (dark) matter to the theory and show that the universe can undergo a `extended period' of matter domination followed by an accelerating dark energy dominated era. This would in principle allow for structure formation in this model. It would be interesting to study such models in more details studying not only the background evolution but also the evolution of perturbation on this background and in particular structure formation. This would eventually allow us to compare such models with experimental data.


\end{document}